\documentclass[12pt]{article}
\usepackage[a4paper, margin=2cm]{geometry}
\usepackage{authblk}

\usepackage[main=english]{babel}
\usepackage{amsmath}
\usepackage{hyperref}       
\usepackage{url}            
\usepackage{amsfonts}       
\usepackage{graphicx}
\usepackage[skip=0pt]{caption}

\title{Orbital magnetic susceptibility of multifold fermions.}

\author{D.A.~Pshenay-Severin, A.T.~Burkov}

\affil{Ioffe Institute, 194021, St-Petersburg, Russia}

\begin{document}
\maketitle

\begin{abstract}
Topological semimetals are intensively studied in recent years. Besides the well known Weyl and Dirac semimetals, some materials possess nodes with linear crossing of multiple bands. Low energy excitations around these nodes are called multifold fermions and can be described by $\mathbf{k}\cdot\mathbf{p}$ Hamiltonian with pseudospin greater than 1/2. In the present work we investigate the contribution of these states into orbital magnetic susceptibility $\chi$. We have found that, similarly to Weyl semimetals, the dependence of susceptibility on chemical potential $\mu$ shows an extremum when $\mu$ is close to the band crossing energy. In the case of half-integer pseudospin, this extremum is a minimum and the susceptibility is negative (diamagnetic). While in the case of integer pseudospin, the susceptibility is large and positive (paramagnetic) due to the contribution of dispersionless band, corresponding to zero pseudospin projection. This leads also to nonmonotonic temperature dependence of $\chi$. As an example, we considered the case of cobalt monosilicide, where the states near the $\Gamma$ point correspond to pseudospin 1 without spin-orbital interaction, and to a combination of Weyl node and pseudospin-3/2 states taking into account spin-orbit coupling.
\end{abstract}

\section{Introduction}
Topological semimetals demonstrate a variety of interesting properties (see, e.g., the review articles  \cite{Yan2017,Wang2017,Burkov2018,Armitage2018,Lv2021}). In the most simple case of Weyl semimetal, electronic spectrum near the Fermi level posses nodes with linear crossing of two bands. The absolute value of topological charge (or Chern number) of a Weyl node is equal to unity. Low energy excitations around such nodes correspond to fermions with (pseudo-)spin 1/2 and can be described by the Hamiltonian

\begin{equation}\label{eq:H_kp}
\hat{H}_J = v\,\hat{\mathbf{J}}\cdot\hat{\mathbf{p}},
\end{equation}
\noindent
where $v$ is a velocity, $\hat{\mathbf{J}}$ is the operator of (pseudo-)spin $J=1/2$, and $\hat{\mathbf{p}}$ is the linear momentum. The Brillouin zone of topological semimetal contains several such nodes so that the total topological charge is equal to zero. The projections of these nodes onto surface Brillouin zone are connected by surface Fermi arcs. Their number is determined by the sum of the absolute values of topological charges of the same sign.

As it was shown in \cite{Bradlyn2016}, materials of certain crystal structures contain nodes with linear crossing of multiple bands. The low energy excitations around these nodes were named unconventional fermions~\cite{Bradlyn2016}. The fermions described by Hamiltonian (\ref{eq:H_kp}) with pseudospin $J>1/2$ belong to this group. An example of such materials is CoSi, crystallizing in cubic noncentrosymmetric structure B20 (space group $P2_13$, No.198)\cite{Tang2017,Pshenay2018}. In CoSi, topological band crossings exist at the time-reversal invariant $\Gamma$ and $R$ points. Their topological charge is 4, and there are four elongated Fermi arcs connecting projections of these points onto the surface Brillouin zone. The peculiarities of the band structure of CoSi were confirmed by ARPES measurements \cite{Takane2019,Rao2019,Sanchez2019}. Without spin-orbit coupling the low-energy excitations near the $\Gamma$ point correspond to fermions with pseudospin $J=1$.
Spin-orbit coupling splits the spectrum into a Weyl node and a node with 4-fold linear band crossing \cite{Tang2017,Pshenay2018}. If one considers spin-orbit coupling in the zeroth order in $\mathbf{k}$ around the $\Gamma$ point, the spectrum of these 4-fold degenerate states can be described by pseudospin-3/2 Hamiltonian. The list of space groups that can host symmetry protected fermions with the pseudospin-3/2 were given in~\cite{Bradlyn2016}. Whether materials with $J\ge2$ can exist is still unknown~\cite{Ezawa2017}. The analysis in~\cite{Bradlyn2016} showed that there are space groups that can host up to eight-fold linear band crossing points but not all of them can be represented by the Hamiltonian of considered form (\ref{eq:H_kp}). For example, it was noted that fermions with pseudospin $J\ge2$ cannot exist at high symmetry points of the Brillouin zone~\cite{Bradlyn2016}.

The nontrivial topology of band structure leads to a number of interesting transport properties in magnetic field, such as chiral anomaly and negative magnetoresistance \cite{Burkov2018,Son2013}. They are associated with the spectrum of Landau levels, which contains the chiral energy level, where carriers can move only along or against magnetic field, depending on the sign of the topological charge. In cobalt monosilicide, these effects were studied in the work \cite{Schnatmann2020}. Quantum oscillations of resistance and thermopower were  studied in CoSi in magnetic fields up to 15~T \cite{Wu2019,Huang2021,Xu2019}.

The magnetic susceptibility of topological semimetals also has many interesting features. The susceptibility of Weyl semimetals was studied in several works reviewed in \cite{Mikitik2019}.
A giant diamagnetic anomaly of orbital susceptibility in Weyl semimetal was predicted in \cite{Mikitik1989}. It is characterized by a logarithmic divergence of susceptibility as the chemical potential approaches the Weyl node energy. This anomaly in Weyl and Dirac semimetals was also studied in \cite{Koshino2010,Koshino2016}. For an arbitrary spectrum slope, it was analyzed in the work \cite{Mikitik2016}. The temperature dependence of susceptibility  turns out to be nonmonotonic and has a minimum at the temperature $T\sim 0.443 \mu/k_B$~\cite{Mikitik2019}, where the chemical potential $\mu$ is measured from the node energy. Interestingly, the minimum of susceptibility was observed in the Weyl semimetal TaAs at approximately 185~K~\cite{Liu2016}. A study of magnetization in TaAs \cite{Zhang2019} also showed that, in contrast to quasiparticles with a nonrelativistic spectrum, there is no saturation of longitudinal magnetization in strong fields in a Weyl semimetal.

Cobalt monosilicide and isostructural materials exhibit a wide variety of magnetic properties. CoSi single crystals at temperatures above 25 K are diamagnetic, but at lower temperatures the magnetic susceptibility becomes paramagnetic in some of the studied samples~\cite{Stishov2012,Narozhnyi2013}. CoSi solid solutions with a small Fe content are diamagnetic at room temperature, however with increasing iron content, ferromagnetic properties appear~\cite{Shinoda1972}.
In the isostructural material MnSi in low magnetic fields, a helical magnetic structure was revealed, which, with increasing field, turns into a ferromagnetic structure~\cite{Stishov2011, Ishikawa1976}. In MnSi, the formation of a skyrmion phase is also possible~\cite{Kanazawa2015}.

The orbital magnetic susceptibility is determined by peculiarities of Landau levels in magnetic field. For topological semimetals, in which a topological charge greater than unity is associated with a nonlinear dependence of energy on the wave vector in some directions, the spectrum in a magnetic field and the density of states were calculated in \cite{Gupta2019}. For fermions with pseudospin 1 and a linear dispersion law, the spectrum in a magnetic field was given in the appendix to the work \cite{Bradlyn2016}. For fermions with higher pseudospin values, Landau levels were calculated in \cite{Ezawa2017}. However, magnetic susceptibility was not considered in these works.

Orbital magnetic susceptibility of fermions with pseudospin 1 was considered in our previous work~\cite{Pshenay2024} and applied to the case of CoSi without spin-orbital coupling. It was shown that the contribution of pseudospin-1 states demonstrate a transition from negative (diamagnetic) to large positive (paramagnetic) susceptibility when the chemical  potential approaches the energy of the node due to the contribution of the flat band states.


In the present work we summarize the results on low-field orbital magnetic susceptibility obtained earlier and extend them to the case of pseudospin greater than unity. As a material example, we study
the susceptibility of multifold fermions in CoSi near the $\Gamma$ point of the Brillouin zone with the account of spin-orbit coupling.


\section{Spectrum of multifold fermions in magnetic field}

\begin{figure}[h]
  \begin{center}
    \includegraphics[width=0.48\textwidth]{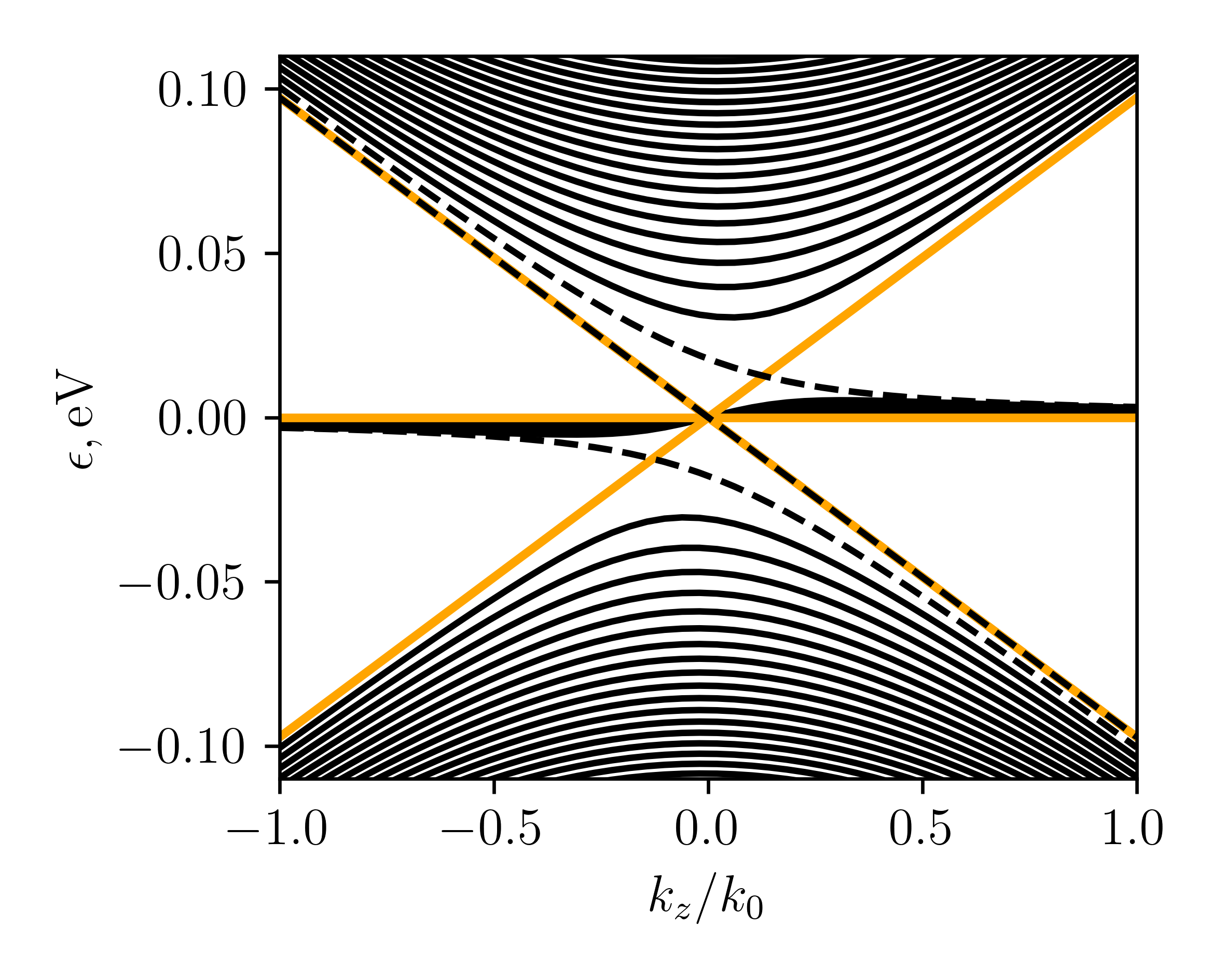}
    \includegraphics[width=0.48\textwidth]{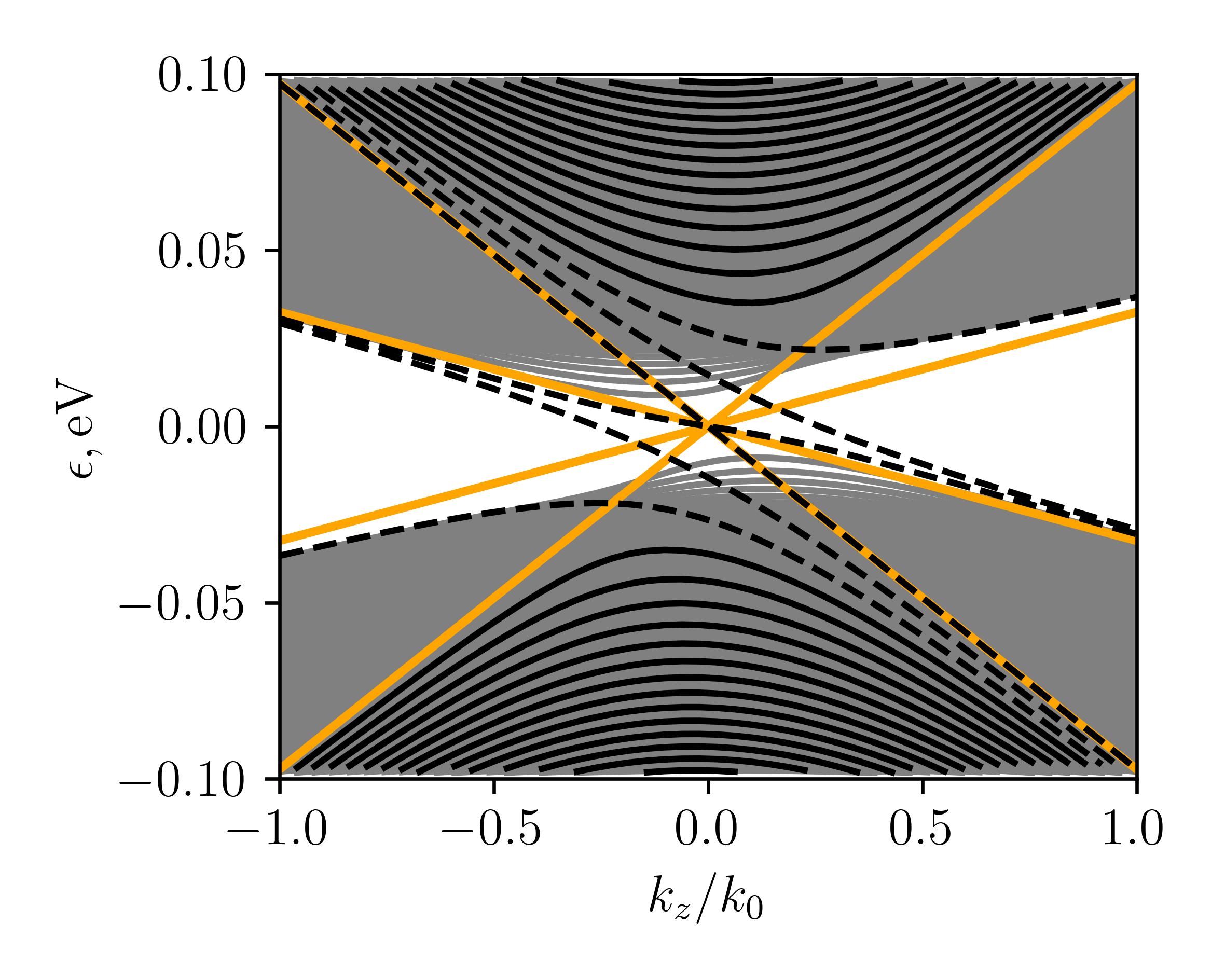}
  \end{center}
  \caption{Zero magnetic field band structure and Landau levels for pseudospin 1 (left panel) and 3/2 (right panel) fermions. Electronic spectrum without magnetic field is plotted with orange lines.
  Landau levels for $B=14$T are plotted with black and gray solid lines for $n\ge0$ and the midgap solutions for $n<0$ are plotted with black dashed lines.
  }\label{fig:LL_1_3_2}
\end{figure}

Let us consider the fermions described by the Hamiltonian (\ref{eq:H_kp}), generalized to the case of $J\ge1/2$. Their electronic spectra in zero magnetic field correspond to a crossing of $2 J+1$ linear bands. For $J=1$ and $3/2$ these spectra are plotted in the figure~\ref{fig:LL_1_3_2} with orange lines. Note that the spectra with integer $J$ contain dispersionless bands, corresponding to zero pseudospin projection.

The spectrum of fermions in magnetic field can be calculated using Pierls substitution. We follow the approach of \cite{Ezawa2017}, slightly reformulated using matrix notation. The momentum operator $\hat{\mathbf{p}}$ is replaced with $\hat{\mathbf{\pi}} = \hat{\mathbf{p}}+e \mathbf{A}/c$, where $\mathbf{A}$ is the vector potential of magnetic field that is equal to $\mathbf{A} = (0, B x, 0)$ in Landau gauge. It is convenient to introduce the rising and lowering operators $\hat{J}_{\pm} = \hat{J}_x \pm i \hat{J}_y$ and $\hat{a}^{+} = (\hat{\pi}_x + i \hat{\pi}_y)/(i \sqrt{2 b} \hbar )$, $\hat{a} = -(\hat{\pi}_x - i \hat{\pi}_y)/(i \sqrt{2 b} \hbar )$. Here, $b = e B/\hbar c$ is proportional to magnetic field and is related to the magnetic length $l$ as $b = l^{-2}$. As a result, the Hamiltonian in magnetic field takes the form
\begin{equation}\label{eq:H_mag}
\hat{H}_J = i v \hbar \sqrt{\frac{b}{2}} \left(\hat{a}^+ \hat{J}_{-} -\hat{a} \hat{J}_{+}\right) + \hbar\,v\,\hat{J}_{z}\,k_z.
\end{equation}

The multicomponent wave function $\psi_j(x)$ ($j=J,J-1,...,-J$) can be taken as a superposition of eigenfunctions of harmonic oscillator $\phi_n(x)$ \cite{Luttinger1955}
\begin{equation}\label{eq:psi_j}
\psi_j(x) = \sum_{n=0}^{\infty}c_{j n} \phi_n(x).
\end{equation}


Multicomponent wave functions of the chiral levels have one or several zero components~\cite{Ezawa2017}. For Weyl fermions, it is $\psi_{1/2}$. In the case of pseudospin-1 fermions, there are two chiral levels and the zero components are either $\psi_{1}$ or both $\psi_{1}$ and $\psi_{0}$. To account for such states in general case, it is convenient to extend low bound of summation in (\ref{eq:psi_j}) down to $n=-(J-j)$, understanding that $c_{j n}=0$ for $n<0$. The substitution of (\ref{eq:psi_j}) into the Schroedinger equation leads to the matrix eigenvalue problem. It can be reduced to closed systems of $2 J+1$ equations for each $n$ that have the form $\hat{\mathcal{H}}\,\mathbf{c} = \epsilon\,\mathbf{c}$, where $\mathbf{c} = \left(c_{J, n}, c_{J-1, n+1}, ..., c_{-J, n+2 J} \right)^{\mathrm{T}}$. The matrix $\hat{\mathcal{H}}$ is tridiagonal with the matrix elements $\mathcal{H}_{j,j}=\hbar\,v\,k_z j$ and $\mathcal{H}_{j,j+1}=\mathcal{H}_{j+1,j}^{*} = i\,\hbar\,v\, (\hat{J}_{+})_{j, j+1} ((n+J-j) \,b/2)^{1/2}$ for $j=J,J-1,...-J$, where $(\hat{J}_{+})_{j, j\pm1} = (J (J+1)-j (j\pm1))^{1/2}$. For each $n\ge0$ one obtains $2 J+1$ subbands of Landau levels $\epsilon_m(n, k_z, b)$, $m=J, J-1, ..., -J$. For $-2 J\le n<0$ first $|n|$ rows and columns of matrix $\hat{\mathcal{H}}$ should be discarded and additional $J (2 J+1)$ midgap solutions appear.

As an example, the spectra of Landau levels for $J=1$ and $3/2$ are plotted in the figure~\ref{fig:LL_1_3_2} for $B=14$T with black and gray lines. In estimations we used $v_0 = 18.6\cdot 10^{6}$cm/s and the maximum wave vector $k_0 = 0.08$\AA$^{-1}$ that are typical for the states with $J=1$ near $\Gamma$ point in CoSi~\cite{Pshenay2018}. To make the range of energies for different $J$ comparable, we used electron velocities equal to $v = v_0/J$.
The Landau levels for $n\ge0$ are plotted with solid lines. The midgap solutions for $n<0$ is convenient to denote by a pair of indexes $(m,\,m')$ of bands that they connect. These solutions are plotted with dashed lines.
For $J=1$ there are three bands of Landau levels and three midgap solutions.
For $J=3/2$ there are four bands of Landau levels and six midgap solutions.

\section{Orbital magnetic susceptibility of multifold fermions}

Magnetic susceptibility $\chi$ is determined by all states filled with electrons, but its dependence on temperature and chemical potential is determined by electronic states with energy close to $\mu$. The influence of topological states will be important if the chemical potential is close to the band crossing energy. Following \cite{Mikitik1989,Koshino2010,Koshino2016}, we split the susceptibility into the contribution from the energy range $\pm\epsilon_{0}$ around the band crossing and a background contribution from the other parts of the electronic spectrum. The contribution of topological states to the low-field magnetic susceptibility of Weyl semimetals of the first and second types were considered in \cite{Mikitik1989,Koshino2010,Koshino2016}. In \cite{Pshenay2024}, the case of $J=1$ was considered. Here we summarize the results obtained earlier and extend them to the case of $J>1$.

Using the obtained electronic spectrum in magnetic field the thermodynamic potential $\Omega$ of the unit volume can be calculated. Then the orbital magnetic susceptibility is defined as $\chi = -\partial^2 \Omega/\partial B^2 = -(e/\hbar c)^2 \partial^2 \Omega/\partial b^2$.

The thermodynamic potential is given as a sum of contributions from the branches of spectrum for $n\ge0$
\begin{equation}\label{eq:Omega_b}
  \Omega_b = -\frac{b \,k_B\, T}{4 \pi^2}
  \sum_{m=J}^{-J}{
    \int_{-k_0}^{k_0}{ dk_z
	  \sum_{n=0}^{n_m}{
        \ln
          \left(
            1 + \exp\left( \frac{\mu-\epsilon_m(n, k_z, b)}{k_B T} \right)
          \right)
      }
    }
    },
\end{equation}
\noindent and the contribution from midgap states for $-2\,J\le n \le -1$
\begin{equation}\label{eq:Omega_m}
  \Omega_m = -\frac{b \,k_B\, T}{4 \pi^2}\sum_{n=-2 J}^{-1}{
  \sum_{m=J+n}^{-J}{
	  \int_{-k_0}^{k_0}{ dk_z
        \ln
          \left(
            1 + \exp\left( \frac{\mu-\epsilon_m(n, k_z, b)}{k_B T} \right)
          \right)
      }
    }
}.
\end{equation}
\noindent In these equations $k_B$ is the Boltzmann constant, $k_0$ determines the range of wave vectors $k_z$ and corresponding energy range $\epsilon_0 = \hbar\,J v\,k_0$.
The equations do not take into account possible spin degeneration.

The upper limit in the sum over $n$ in Eq. (\ref{eq:Omega_b}) is determined by the restriction on electronic energies $|\epsilon_m(n_m, k_z, b)|\le\epsilon_0$. In the low-field limit the summation over Landau levels $n$ can be replaced with the integration, using Euler–Maclaurin formula \cite{Shoenberg2009}. Let us introduce a variable $x_m$ such that $|\epsilon_m(x_m, k_z, b)|=\epsilon_0$ then $n_m=[x_m]$, where square brackets denote an integer part. For a function $\mathcal{G}(n)$ the sum can be approximately calculated as
\begin{equation}\label{eq:EM_1}
\begin{split}
\sum_{n=0}^{n_m} \mathcal{G}(n) &\approx\int_0^{x_m}{\mathcal{G}(x) dx}+\frac{1}{2}\mathcal{G}(0)-\frac{1}{12}\mathcal{G}'(0) \\
& +\left(\frac{1}{2}-\delta \right) \mathcal{G}(x_m)
+\left(\frac{1}{12} -\frac{\delta }{2}
+ \frac{\delta^2}{2}\right)\mathcal{G}'(x_m),
\end{split}
\end{equation}
\noindent where $\delta = x_m - n_m$.
When magnetic field increases, energy levels leave the range $|\epsilon|\le\epsilon_0$ that leads to oscillations of $\Omega$ with inverse magnetic field due to the change of $\delta$. These non-physical oscillations appeared because we consider finite energy range around chemical potential. They can be eliminated in the low-field limit using averaging over inverse magnetic field $1/B$ when $x_m$ changes from $n_m$ to $n_m+1$. This averaging leads to $\langle \delta \rangle = 1/2$ and $\langle \delta^2 \rangle = 1/3$ if the derivative $\partial x_m/\partial(B^{-1})$ does not depend on magnetic field. This derivative is a constant for parabolic dispersion and for $J\le1$. For larger $J$ it approaches constant value at large $n_m$, e.g. in the low-field limit.

Substituting the powers of $\delta$ with corresponding average values in (\ref{eq:EM_1}), gives
\begin{equation}\label{eq:EM}
\sum_{n=0}^{n_m} \mathcal{G}(n) \approx\int_0^{x_m}{\mathcal{G}(x) dx}+\frac{1}{2}\mathcal{G}(0)-\frac{1}{12}\mathcal{G}'(0).
\end{equation}

Using this equation allows one to evaluate orbital susceptibility analytically for the case of $J\le3/2$, when required derivatives can be calculated using the secular equation.

In the Weyl semimetal, the chiral level does not depend on magnetic field and does not contribute to susceptibility, which in this case has the form \cite{Mikitik2019}
\begin{align}\label{eq:susc_1_2}
  \chi_{1/2} =& -\frac{v}{24 \pi^2 \hbar} \left( \frac{e}{c} \right)^2
    \int_{0}^{\epsilon_0}{ \frac{d\epsilon}{\epsilon}
      \left( f_0(-\epsilon) - f_0(\epsilon) \right)
      },
\end{align}
\noindent where $f_0(\epsilon) = 1/\left(1+\exp[(\epsilon-\mu)/(k_B T)]\right)$ is the Fermi distribution. Note, that compared to \cite{Mikitik2019, Pshenay2024}, the factor of $1/4$ was introduced in Eq.~(\ref{eq:susc_1_2}). This is due to the fact that in Refs.~\cite{Mikitik2019, Pshenay2024} Pauli matrices were used instead of pseudospin operators in the Hamiltonian (\ref{eq:H_kp}) and also because we consider bands without spin degeneracy.
At zero temperature, susceptibility (\ref{eq:susc_1_2}) diverges logarithmically when the chemical potential approaches band crossing energy \cite{Mikitik2019}.

For $J=1$, there are three midgap solutions (fig.~\ref{fig:LL_1_3_2}, left panel) that can be denoted by a pair of indexes $(m,\,m')$ of the bands that they connect. The first midgap solution $(-1, 1)$ does not depend on magnetic field, but others ($(1,0)$, $(0,-1)$) do. It is convenient to combine together the contributions of bands $m=\pm1$, the midgap solution $(-1, 1)$ and two halves of midgap solutions $(1,0)$ for $k_z<0$ and $(0,-1)$ for $k_z>0$. This contribution has the same form as Eq.~(\ref{eq:susc_1_2}) for the Weyl node \cite{Pshenay2024}, if one replaces $v$ by $2 v$, so that the slopes of the bands were the same.

In the case of pseudospin $J=1$, there is also a contribution from dispersionless band ($m=0$) and midgap solutions $(1,0)$ for $k_z>0$ and $(0,-1)$ for $k_z<0$. They lie in the bounded energy interval that leads to high density of states. In this case it is not possible to determine $n_m$ from the limiting energy $\epsilon_0$. The number of levels in this group can be obtained from the conservation of the total number of states.
Without magnetic field, the number of states in the dispersionless band in the sphere of the radius $k_0$ is equal to $\mathcal{N} = k_0^3/(6\,\pi^2)$. In small magnetic fields, the number of states can be calculated using equation $b k_0 (\sum_{n=0}^{n_m} \mathcal{G}(n) + 1) / (2 \pi^2) \approx b k_0 (x_m+3/2)/(2 \pi^2)$, where $\mathcal{G}(n)=1$ and the unity in the brackets accounts for additional midgap state for each $k_z$. The sum over $n$ was calculated using Eq.~(\ref{eq:EM}). Equating this result to $\mathcal{N}$, one obtains $x_m = k_0^2/(3\,b)-3/2$. Then the contribution from these bands to the susceptibility is equal to
\begin{align}\label{eq:susc_1_0}
  \chi_{1,0}
&= \frac{v^2 k_0}{4\pi^2}\,\frac{\arctan\left(\sqrt{3/2}\right)}{\sqrt{3/2}} \left( \frac{e}{c} \right)^2 \left(- \frac{\partial f_0}{\partial \epsilon} \right)_{\epsilon=0}.
\end{align}

\noindent Interestingly, this contribution is positive (paramagnetic) in contrast to diamagnetic contribution of $\chi_{1,\pm1}$. Both contributions have extrema near $\mu=0$, but due to different magnitudes and dependences on chemical potential, the total susceptibility $\chi_1$ becomes positive when $\mu$ is close to zero.

In the case of pseudospin $J=3/2$, similar calculations lead to
\begin{align}\label{eq:susc_32}
  \chi_{3/2} =& -\frac{v}{4 \pi^2 \hbar} \left( \frac{e}{c} \right)^2
    \int_{0}^{\hbar v k_0}{\mathcal{F}(\epsilon) d\epsilon},
\end{align}
\noindent where
\begin{align*}
\mathcal{F}(\epsilon) =& \frac{9}{16}
\int_{\epsilon/2}^{3\epsilon/2}{
\frac{4 u^2 - 9 \epsilon^2}{u^4}
(f_0(-u) - f_0(u)) du
} \\
+& \frac{1}{6 \epsilon}
\left(
73 \left(f_0\left(-\frac{\epsilon}{2}\right)-f_0\left(\frac{\epsilon}{2}\right)\right)
+
3 \left(f_0\left(-\frac{3\epsilon}{2}\right)-f_0\left(\frac{3\epsilon}{2}\right)\right)
\right).
\end{align*}

\begin{figure}[h]
  \begin{center}
    \includegraphics[width=80mm]{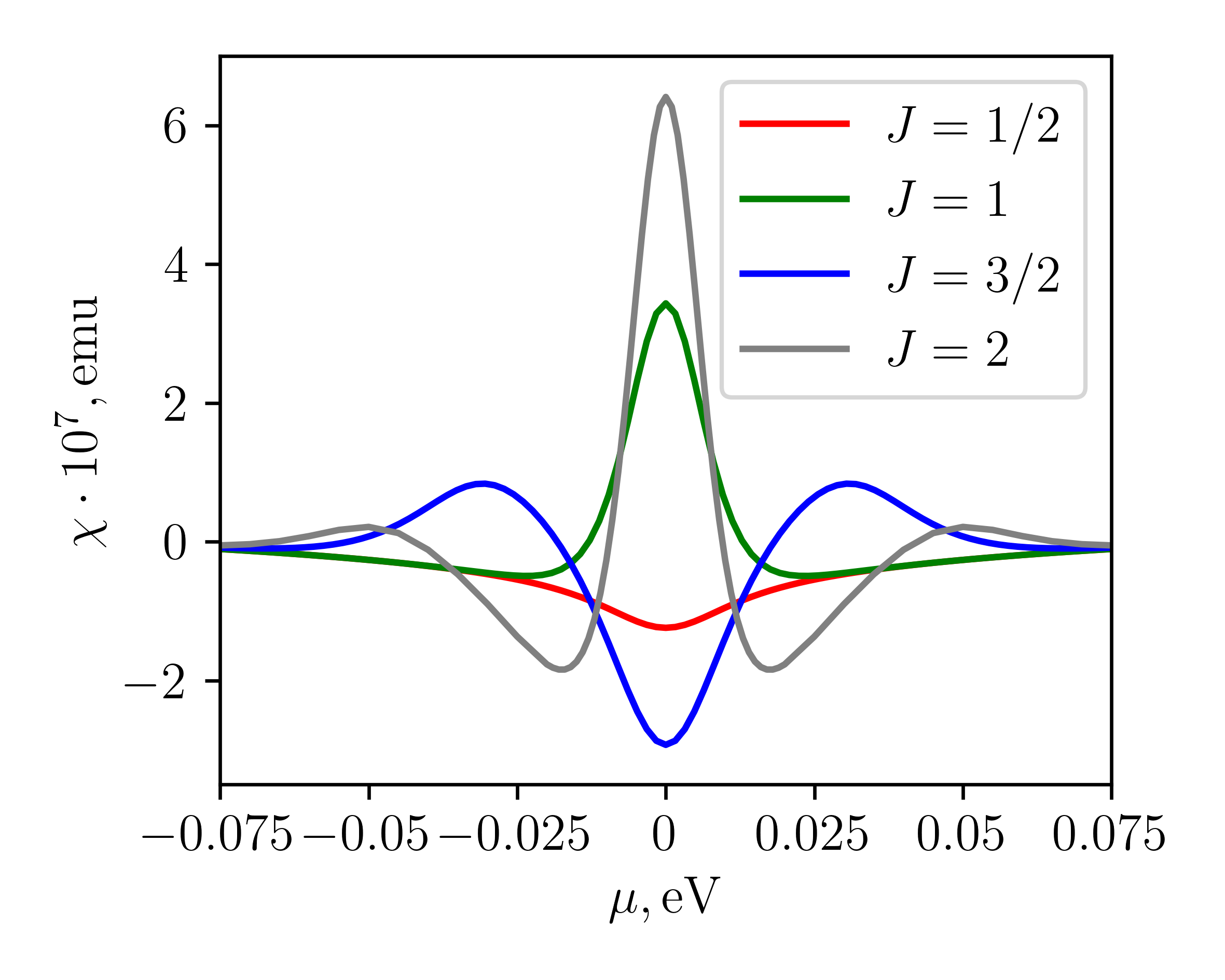}
    \includegraphics[width=80mm]{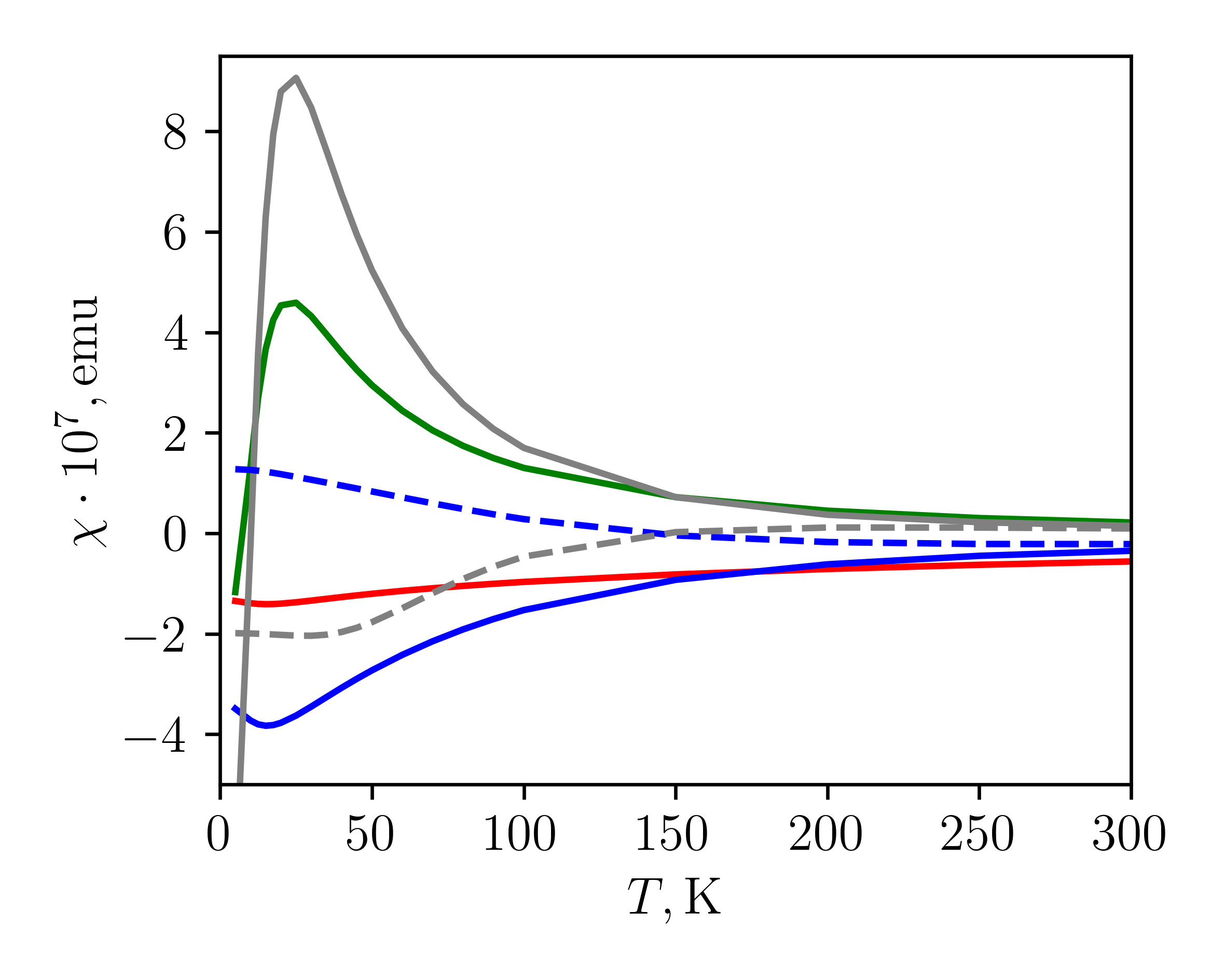}
  \end{center}
  \caption{Low-field orbital magnetic susceptibility of fermions with different pseudospins $J$ as a function of chemical potential $\mu$ at 50K (left panel) and as a function of temperature (right panel). The curves for pseudospins $J=$1/2, 1, 3/2 and 2 are plotted with red, green, blue and gray colors on both panels. Solid curves in the right panel are plotted for $\mu=-0.003$eV. Dashed blue curve is plotted for $J=3/2$ and $\mu=0.03$eV, and dashed gray is plotted for $J=2$ and $\mu=0.02$eV, corresponding to secondary extrema in the left panel.}\label{fig:mag_susc_pse}
\end{figure}

For the case of higher pseudospin, we calculated low-field susceptibility numerically. The smoothed thermodynamic potential was calculated using Euler–Maclaurin formula (\ref{eq:EM}) and its field dependence was fitted by a second order polynomial in the low-field range below 1T. For $J=2$, corresponding dependence of susceptibility on the chemical potential at 50K is plotted in the left panel of the figure \ref{fig:mag_susc_pse}. The temperature dependences for several chemical potential values are plotted in the right panel of the figure \ref{fig:mag_susc_pse}.

If electron velocity $v$ and wave vector range $k_0$ are the same for different pseudospins, then the susceptibility increases with $J$ due to an increase of the number of energy branches and considered energy range. To facilitate the comparison, the susceptibility was calculated for scaled velocity $v = v_0/J$ so that the energy range $\epsilon_0$ was the same. Even in this case the magnitudes of extrema and the number of extrema in the dependence of $\chi(\mu)$  increases with $J$. For half-integer pseudospins the main extremum is a minimum and for integer pseudospins it is a maximum due to the contribution of bounded middle bands. The temperature dependences of $\chi$ for integer pseudospins also exhibit maxima below 50K. In general, when chemical potential shifts from the energy of band-crossing point, $\chi$ becomes smaller. If the chemical potential reaches the next extremum, the susceptibility can still be quite large, but of the opposite sign, as shown by dashed lines in the right panel of the figure \ref{fig:mag_susc_pse} for $J=3/2$ and 2.

Large values and non-monotonic dependence of orbital susceptibility on temperature and chemical potential were observed in several materials with linear energy dispersion. For example, the minimum in the temperature dependence of magnetic susceptibility was observed in Weyl semimetal TaAs\cite{Liu2016} at about 185K. Large values of diamagnetic susceptibility in graphene were discussed in \cite{McClure1956} and were explained by the contribution of virtual interband transitions.

Another unusual feature, the change of the sign of orbital susceptibility, was also discussed in the literature. For example, the change of the sign of orbital susceptibility was considered in several two-dimensional lattices. In \cite{Vignale1991} it was shown that orbital susceptibility is always positive (paramagnetic) if the chemical potential is close to a saddle point of the band structure. The change of the sign of $\chi$ from negative to positive during transition from graphene to the dice lattice was considered in \cite{Raoux2014}.

\section{The contribution of topological states near the $\Gamma$ point to orbital susceptibility of CoSi}

As an application to real material, we calculated the contribution to low-field magnetic susceptibility from the states near the $\Gamma$ point in CoSi. Without spin-orbit coupling these states can be described by the Hamiltonian $\hat{H}_{J=1}$ (\ref{eq:H_kp}) for the pseudospin $J=1$.
In order to account for spin-orbit coupling, it is necessary to expand the basis set to include spin $\vert J=1,j \rangle \otimes \vert \uparrow \rangle, \vert J=1,j \rangle \otimes \vert \downarrow \rangle, j=1,0,-1$. The calculations showed that spin-orbit coupling in the zeroth order with respect to $\mathbf{k}$ can be written as $\hat{H}^{SOC} = 2\,\Delta\,\hat{\mathbf{J}}\cdot\hat{\mathbf{s}}$, where $\hat{\mathbf{s}}$ is the operator of electron spin and the parameter $\Delta$ determines the magnitude of splitting. Thus, near the $\Gamma$ point the Hamiltonian has the form $\hat{H} = \hat{H}_{J=1}\otimes \hat{1}_{2\times 2} + \hat{H}^{SOC}$. Zeeman splitting can be taken into account by additional term $\hat{H}^{(Z)} = \mu_B B  \hat{1}_{3\times 3}\otimes \hat{\sigma}_{3}$, where $\mu_B$ - Bohr magneton and $\hat{\sigma}_{3}$ - Pauli matrix.

\begin{figure}[h]
  \begin{center}
    \includegraphics[width=0.49\textwidth]{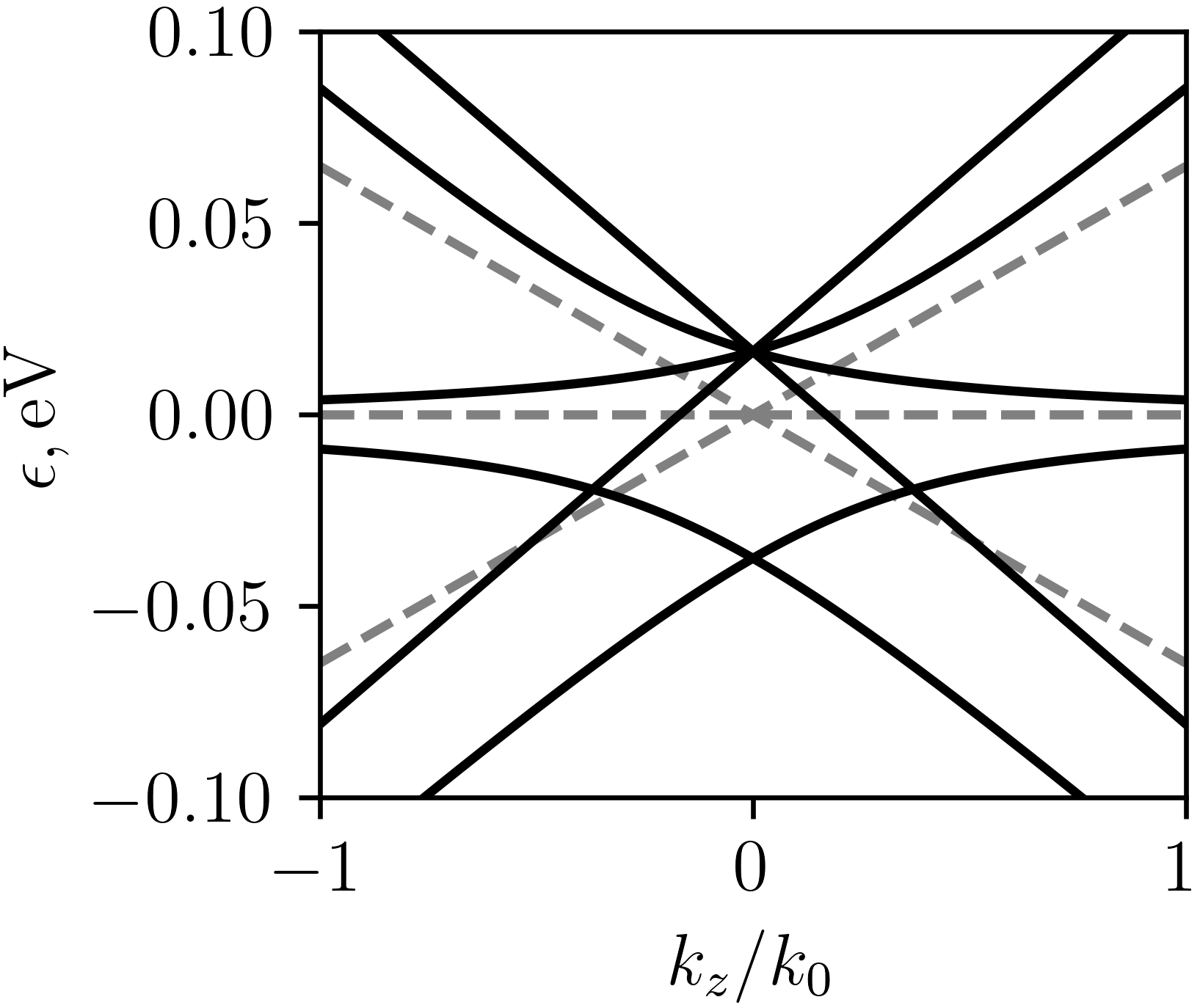}  \includegraphics[width=0.49\textwidth]{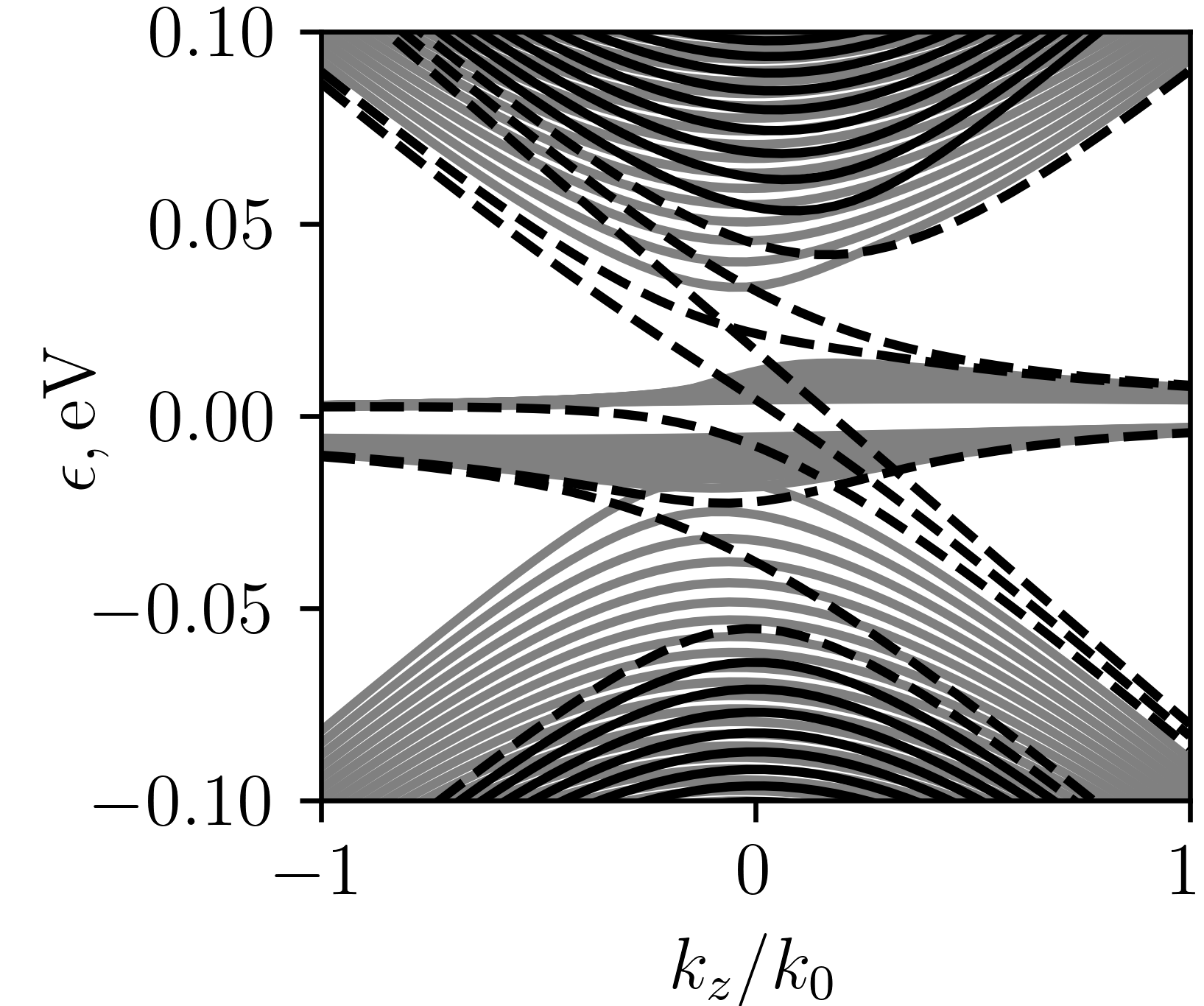}
  \end{center}
  \caption{Left panel: Electronic structure of CoSi near the $\Gamma$ point in zero magnetic field. The spectrum with (without) the account of SOC is plotted with black (gray dashed) lines.
  Right panel: Landau levels with the account of spin-orbit splitting at $B=14$T. Energy levels with $n\ge0$ ($n<0$) are plotted with solid (dashed) lines.}
  \label{fig:LL_CoSi_SOC}
\end{figure}

The electronic spectrum of CoSi near the $\Gamma$ point in zero magnetic field is plotted in the left panel of the figure~\ref{fig:LL_CoSi_SOC}. Gray dashed lines show the spectrum without SOC. These spectrum lines are doubly degenerate due to spin. With the account of SOC, the 6-fold degenerate level at the $\Gamma$ point splits into a 4-fold degenerate level ($J=3/2$) and a doublet ($J=1/2$), shifted in energy by $\Delta$ and $-2\Delta$, respectively. They are plotted in the panel with the black lines. First-principle calculations yield $\Delta = 18$~meV\cite{Pshenay2018}.

Without SOC, the contribution of these states into magnetic susceptibility was calculated in our previous work \cite{Pshenay2024}. Landau levels for this case are plotted in the left panel of the figure~\ref{fig:LL_1_3_2}. The dependence of susceptibility on chemical potential and temperature is similar to that shown in the figure \ref{fig:mag_susc_pse} for $J=1$ and differ only by a factor of two due to spin degeneracy.

Taking into account SOC, electronic spectrum in magnetic field can be obtained for any $J$ similarly to described above.
 $\hat{\mathcal{H}}$ will be of five-diagonal form with matrix elements
$\mathcal{H}_{j\sigma; j \sigma}=\hbar\,v\,k_z j + \mu_B\,B + 2\,\Delta\,j\,\sigma$,
$\mathcal{H}_{j\sigma; j+1,\,\sigma} = \mathcal{H}^{*}_{j+1,\,\sigma; j,\,\sigma}
= i\,\hbar\,v\,(\hat{J}_{+})_{j, j+1} ((n+\nu_{j,\,\sigma}) \,b/2)^{1/2}$,
and
$\mathcal{H}_{j\sigma; j+1,\,\sigma-1} = \mathcal{H}^{*}_{j+1,\,\sigma-1; j,\,\sigma}
= \Delta\,(\hat{J}_{+})_{j, j+1} (\hat{s}_{-})_{\sigma, \sigma-1}$, where $\nu_{j,\,\sigma} = J-j+s-\sigma$
and $(\hat{s}_{\pm})_{\sigma, \sigma\pm1} = (s (s+1) - \sigma (\sigma\pm1))^{1/2}$ are matrix elements of spin rising and lowering operators. Here, $s=1/2$ is the spin of electron, and $\sigma=1/2, -1/2$ enumerates its projections on $z$-axis.
For given $n\ge0$, the solution of eigenvalue problem for $\hat{\mathcal{H}}$ gives $2 (2 J+1)$ energy levels $\epsilon_{m,\,\gamma}(n, k_z, b)$ and corresponding sets of expansion coefficients
for $(m,\,\gamma)=(J,1/2),\,(J,-1/2), ..., (-J,1/2),\,(-J,-1/2)$. Additional midgap solutions can be obtained for each $n=-(2 J+1), ..., -1$.
Similarly to above consideration, the coefficients $c_{j,\,\sigma;\, n+\nu_{j,\,\sigma}} = 0$, when $n+\nu_{j,\,\sigma}<0$.

\begin{figure}[h]
  \begin{center}
    \includegraphics[width=120mm]{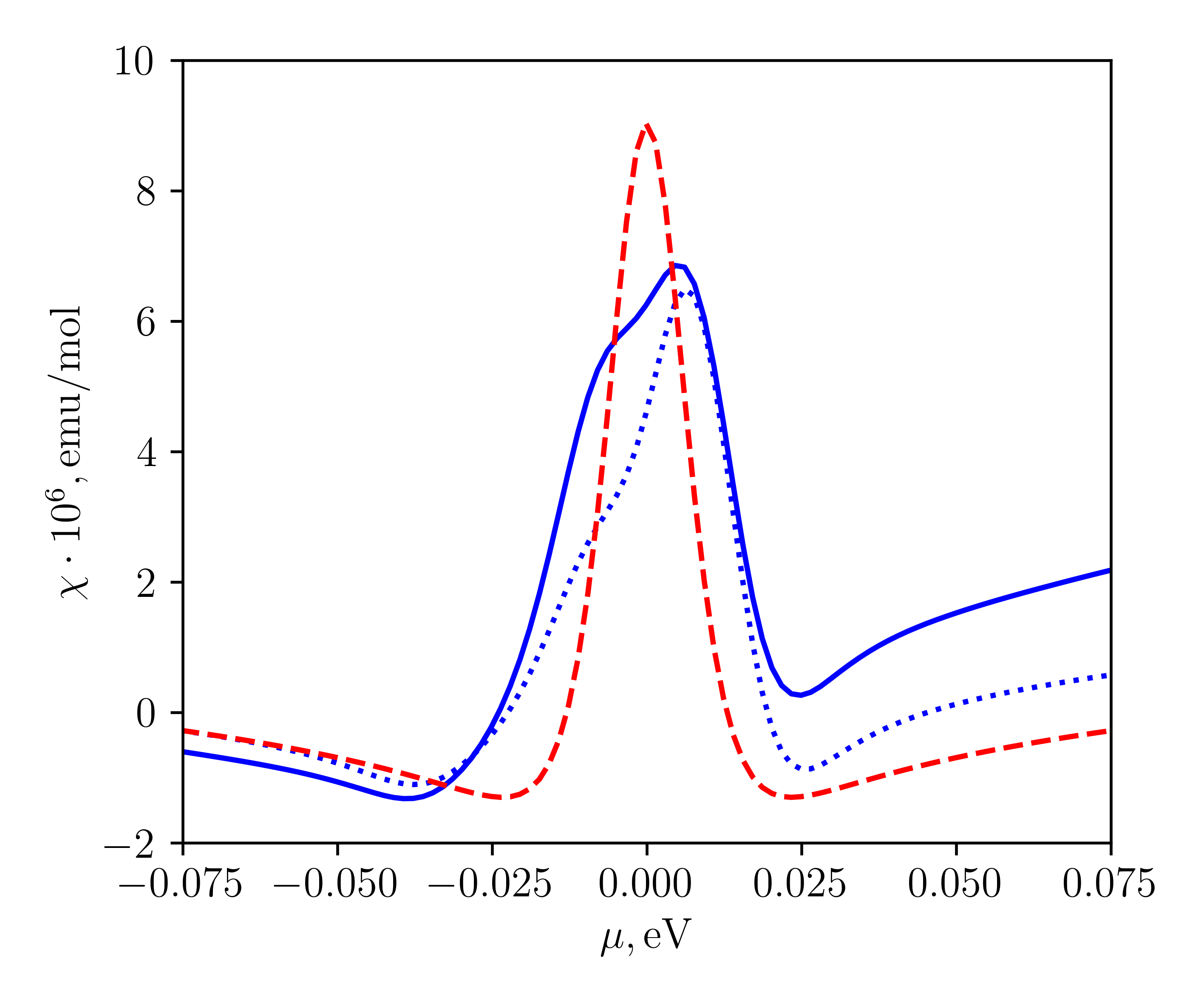}
  \end{center}
  \caption{The contribution of states near the $\Gamma$ point of CoSi to low-field magnetic susceptibility as a function of chemical potential $\mu$ at 50K (solid blue curve). Blue dotted curve is plotted for the case without Zeeman spin splitting in magnetic field, and red dashed curve is the plot for $J=1$ fermions (without SOC) multiplied by 2 due to spin degeneracy.}\label{fig:mag_susc_CoSi}
\end{figure}

\begin{figure}[h]
  \begin{center}
    \includegraphics[width=120mm]{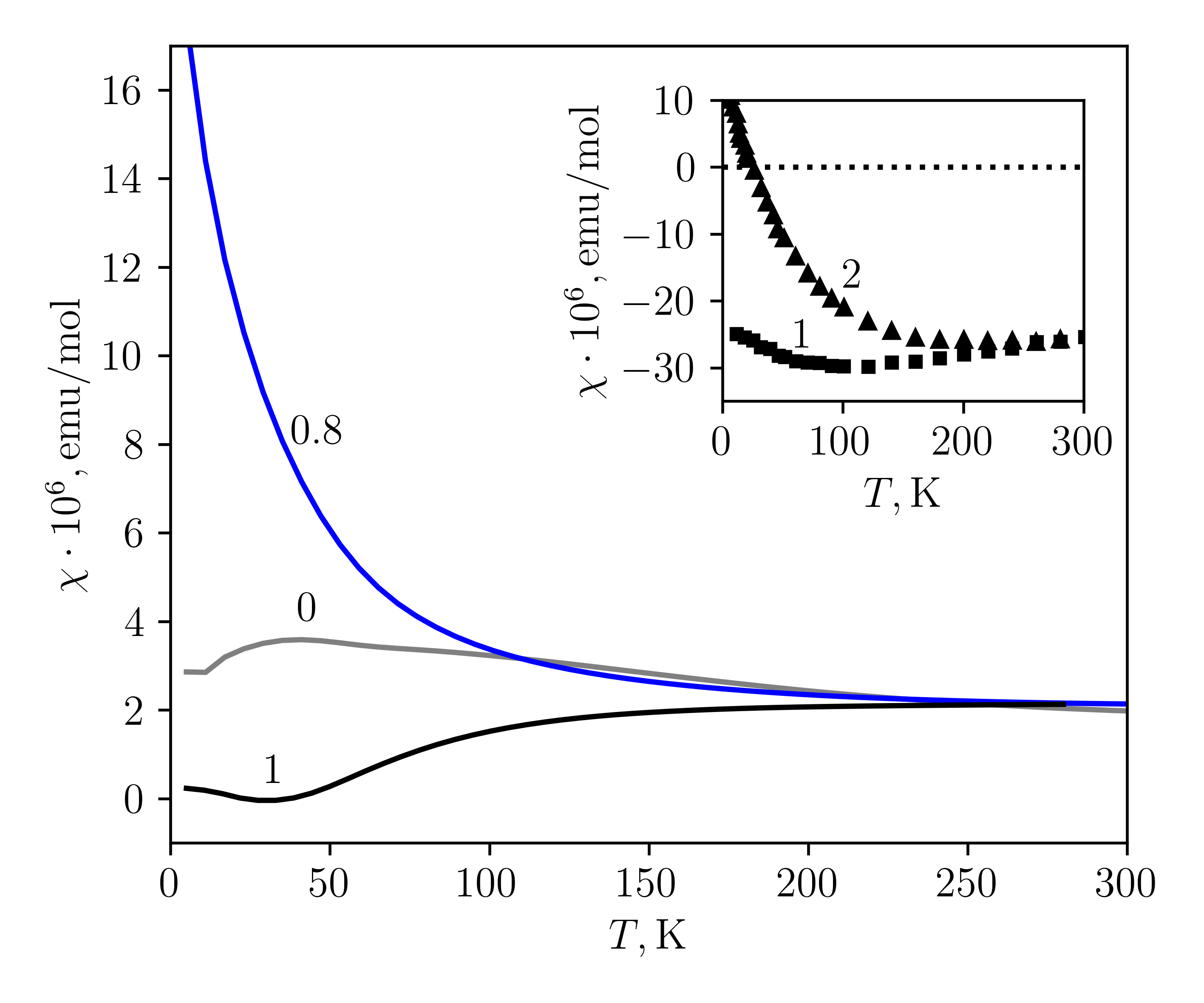}
  \end{center}
  \caption{The contribution of states near the $\Gamma$ point of CoSi to low-field magnetic susceptibility as a function of temperature. Numbers close to curves denote Ni content in atomic percents. Gray curve (0) is plotted for stoichiometric CoSi. Blue (black) curves are plotted for the case of 0.8 (1)~at.\% of Ni doping. Inset shows experimental data from \cite{Stishov2012} for two CoSi samples with residual resistivity ratio $\rho(294\mathrm{K})/\rho(2\mathrm{K}) = $ 2.2 (sample 1) and 6.8 (sample 2).}\label{fig:mag_susc_CoSi_T}
\end{figure}

The Landau levels for CoSi with the account of spin-orbit splitting are plotted in the right panel of the figure~\ref{fig:LL_CoSi_SOC}. The band structure  resembles the combination of Landau levels for $J=3/2$ and $1/2$-fermions shifted by $\Delta$ and $-2\Delta$ and distorted due to SOC.
If the spin-orbit splitting $\Delta$ tends to zero, the band structure will have the form shown in the figure~\ref{fig:LL_1_3_2} (left panel) for pseudospin $J=1$.

The contribution of the states near the $\Gamma$ point into low-field magnetic susceptibility with the account of SOC is plotted with the blue solid curve in the figure~\ref{fig:mag_susc_CoSi} as a function of the chemical potential.
The chemical potential in the figure~\ref{fig:mag_susc_CoSi} is measured from the band crossing energy without SOC.
For comparison, the same calculation without Zeeman splitting is plotted with blue dotted curve. It corresponds to the contribution of the orbital part. The susceptibility without SOC and Zeeman splitting (doubled due to spin degeneracy) is plotted with the red dashed curve. It can be seen that the magnitudes of $\chi$ are similar for all considered cases. In the case of finite SOC strength, there are two bands with the bounded spectrum corresponding to two overlapping peaks in the figure.

The temperature dependence of susceptibility, taking into account the variation of the chemical potential at fixed doping level, is plotted in the figure~\ref{fig:mag_susc_CoSi_T}.
In stoichiometric CoSi, the chemical potential at low temperature is about -13~meV. It is situated between the Weyl point at the energy $-2\Delta=-36$meV and the four-fold band crossing at $\Delta=$~18~meV.
When the temperature vary from 5K to 300K, the chemical potential increases from $\mu=$-13~meV to 3~meV.
Thus, at 5K it lies below the maximum of $\chi$ in the figure~\ref{fig:mag_susc_CoSi} and the susceptibility has a moderate value.
As the temperature increases, the chemical potential shifts towards the maximum susceptibility, but  the magnitude of susceptibility at the maximum decreases with temperature. As a result, the dependence of susceptibility on temperature has a small maximum (gray curve in the figure~\ref{fig:mag_susc_CoSi_T}).

The chemical potential can be increased, for example, by means of Ni doping.
At 0.8~at.\% of Ni doping (additional 0.032 $\overline{e}$ per unit cell), the chemical potential at low temperatures corresponds to the maximum of susceptibility on the blue curve in the figure~\ref{fig:mag_susc_CoSi}. The temperature dependence of $\chi$ for this case is plotted with the blue curve in the figure~\ref{fig:mag_susc_CoSi_T}. It demonstrates the largest temperature variation of the order of $10^{-5}$emu/mol.

At higher Ni doping (1~at.\% of Ni or additional 0.04 $\overline{e}$ per unit cell), the chemical potential at low temperature approaches four-fold band crossing point and the susceptibility decreases because of the negative contribution to $\chi$ from these states. The temperature dependence for this case is plotted with the black curve in the figure~\ref{fig:mag_susc_CoSi_T}.

The measurements of magnetic susceptibility in CoSi~\cite{Stishov2012} showed that at high temperatures it is diamagnetic (negative) (see inset in the figure \ref{fig:mag_susc_CoSi_T}). For less pure sample (1) with smaller residual resistivity ratio, the sign of susceptibility at low temperatures remains negative. But for the sample (2) with higher residual resistivity ratio, $\chi$ change to positive (paramagnetic) values below 25K~\cite{Stishov2012}.
The maximum positive values of $\chi$ were about $10^{-5}$emu/mol. The differences in the temperature dependences of susceptibility for different samples in Ref.~\cite{Stishov2012} may be due to different deviations from stoichiometry.

In our calculations we took into account the contribution to $\chi$ from the states near the $\Gamma$ point, located close to the Fermi level. As was discussed at the beginning of the Section 3, this accounts for the change of susceptibility with the temperature and chemical potential, but does not give its total value.
The calculated temperature dependences in the figure~\ref{fig:mag_susc_CoSi_T}~show that, at certain electron concentration, the variation of $\chi$ with the temperature can be similar in magnitude to measured experimentally.
This suggests that the observed dependence of magnetic susceptibility on composition and temperature~\cite{Stishov2012} may, at least in part, be due to the contribution of topological states.
It would be interesting in the future to consider the contribution to $\chi$ from the other parts of the spectrum to verify this assumption.

\section{Conclusion}

In this work we calculated low-field orbital magnetic susceptibility $\chi$ for multifold fermions with pseudospin $J>1$. Its dependence on temperature $T$ and chemical potential $\mu$ was considered. It was shown that $\chi(\mu)$ is nonmonotonic and demonstrate several extrema when $\mu$ approaches the energy of band crossing. The number of extrema increases with $J$. When $J$ is half-integer, the main extremum is diamagnetic (negative) and it is a minimum, while for integer pseudospin $\chi$ is paramagnetic (positive) and corresponds to a maximum. The temperature dependence of susceptibility also shows nonmonotonic behaviour. The calculations were extended to include spin-orbital coupling and were illustrated by considering the contribution to susceptibility from multifold fermions near the $\Gamma$ point in CoSi.

\section{Acknowledgements}

We are grateful to Dr. Y.V. Ivanov for valuable comments and stimulating discussions.

\bibliographystyle{unsrt}
\bibliography{bib}

\begin{thebibliography}{10}

\bibitem{Yan2017}
B.~Yan and C.~Felser.
\newblock Topological materials: {Weyl} semimetals.
\newblock {\em Annual Review of Condensed Matter Physics}, 8:337--354, 2017.

\bibitem{Wang2017}
Sh. Wang, B.-Ch. Lin, A.-Q. Wang, D.-P. Yu, and Zh.-M. Liao.
\newblock Quantum transport in dirac and {Weyl} semimetals: a review.
\newblock {\em Advances in Physics: X}, 2(3):518--544, 2017.

\bibitem{Burkov2018}
A.A. Burkov.
\newblock {Weyl} metals.
\newblock {\em Annual Review of Condensed Matter Physics}, 9(1):359--378, 2018.

\bibitem{Armitage2018}
N.~P. Armitage, E.~J. Mele, and A.~Vishwanath.
\newblock {Weyl} and dirac semimetals in three-dimensional solids.
\newblock {\em Rev. Mod. Phys.}, 90:015001, 2018.

\bibitem{Lv2021}
B.~Q. Lv, T.~Qian, and H.~Ding.
\newblock Experimental perspective on three-dimensional topological semimetals.
\newblock {\em Rev. Mod. Phys.}, 93:025002, 2021.

\bibitem{Bradlyn2016}
B.~Bradlyn, J.~Cano, Zh. Wang, M.~G. Vergniory, C.~Felser, R.~J. Cava, and B.A.
  Bernevig.
\newblock Beyond dirac and {Weyl} fermions: Unconventional quasiparticles in
  conventional crystals.
\newblock {\em Science}, 353(6299):aaf5037, 2016.

\bibitem{Tang2017}
P.~Tang, Q.~Zhou, and Sh.-Ch. Zhang.
\newblock Multiple types of topological fermions in transition metal silicides.
\newblock {\em Phys. Rev. Lett.}, 119:206402, 2017.

\bibitem{Pshenay2018}
D.A. Pshenay-Severin, Y.V. Ivanov, A.A. Burkov, and A.T. Burkov.
\newblock Band structure and unconventional electronic topology of {CoSi}.
\newblock {\em Journal of Physics: Condensed Matter}, 30(13):135501, 2018.

\bibitem{Takane2019}
D.~Takane, Zh. Wang, S.~Souma, K.~Nakayama, T.~Nakamura, H.~Oinuma, Y.~Nakata,
  H.~Iwasawa, C.~Cacho, T.~Kim, K.~Horiba, H.~Kumigashira, T.~Takahashi,
  Y.~Ando, and T.~Sato.
\newblock Observation of chiral fermions with a large topological charge and
  associated fermi-arc surface states in {CoSi}.
\newblock {\em Phys. Rev. Lett.}, 122:076402, 2019.

\bibitem{Rao2019}
Zh. Rao, H.~Li, T.~Zhang, Sh. Tian, Ch. Li, B.~Fu, C.~Tang, L.~Wang, Zh. Li,
  W.~Fan, J.~Li, Y.~Huang, Zh. Liu, Y.~Long, Ch. Fang, H.~Weng, Y.~Shi, H.~Lei,
  Y.~Sun, T.~Qian, and H.~Ding.
\newblock Observation of unconventional chiral fermions with long fermi arcs in
  {CoSi}.
\newblock {\em Nature}, 567:496--499, 2019.

\bibitem{Sanchez2019}
D.S. Sanchez, I.~Belopolski, T.A. Cochran, X.~Xu, J.-X. Yin, G.~Chang, W.~Xie,
  K.~Manna, V.~Süß, Ch.-Y. Huang, N.~Alidoust, D.~Multer, S.S. Zhang,
  N.~Shumiya, X.~Wang, G.-Q. Wang, T.-R. Chang, C.~Felser, S.-Y. Xu, Sh. Jia,
  H.~Lin, and M.Z. Hasan.
\newblock Topological chiral crystals with helicoid-arc quantum states.
\newblock {\em Nature}, 567:500--505, 2019.

\bibitem{Ezawa2017}
M.~Ezawa.
\newblock Chiral anomaly enhancement and photoirradiation effects in multiband
  touching fermion systems.
\newblock {\em Phys. Rev. B}, 95:205201, 2017.

\bibitem{Son2013}
D.~T. Son and B.~Z. Spivak.
\newblock Chiral anomaly and classical negative magnetoresistance of {Weyl}
  metals.
\newblock {\em Phys. Rev. B}, 88:104412, 2013.

\bibitem{Schnatmann2020}
L.~Schnatmann, K.~Geishendorf, M.~Lammel, C.~Damm, S.~Novikov, A.~Thomas,
  A.~Burkov, H.~Reith, K.~Nielsch, and G.~Schierning.
\newblock Signatures of a charge density wave phase and the chiral anomaly in
  the fermionic material cobalt monosilicide {CoSi}.
\newblock {\em Advanced Electronic Materials}, 6(2):1900857, 2020.

\bibitem{Wu2019}
D.~S. Wu, Z.~Y. Mi, Y.~J. Li, W.~Wu, P.~L. Li, Y.~T. Song, G.~T. Liu, G.~Li,
  and J.~L. Luo.
\newblock Single crystal growth and magnetoresistivity of topological semimetal
  {CoSi}.
\newblock {\em Chinese Physics Letters}, 36(7):077102, 2019.

\bibitem{Huang2021}
X.~Huang, Ch. Guo, C.~Putzke, J.~Diaz, K.~Manna, Ch. Shekhar, C.~Felser, and
  P.J.W. Moll.
\newblock {Non-linear Shubnikov-de Haas oscillations in the self-heating
  regime}.
\newblock {\em Applied Physics Letters}, 119(22), 2021.

\bibitem{Xu2019}
X.~Xu, X.~Wang, T.A. Cochran, D.S. Sanchez, G.~Chang, I.~Belopolski, G.~Wang,
  Y.~Liu, H.-J. Tien, X.~Gui, W.~Xie, M.Z. Hasan, T.-R. Chang, and Sh. Jia.
\newblock Crystal growth and quantum oscillations in the topological chiral
  semimetal {CoSi}.
\newblock {\em Physical Review B}, 100:045104, 2019.

\bibitem{Mikitik2019}
G.~P. Mikitik and Yu.~V. Sharlai.
\newblock Magnetic susceptibility of topological semimetals.
\newblock {\em Journal of Low Temperature Physics}, 197(3):272--309, 2019.

\bibitem{Mikitik1989}
G.P. Mikitik and I.V. Svechkarev.
\newblock Giant anomalies of magnetic susceptibility due to energy band
  degeneracy in crystals.
\newblock {\em Sov. Journal of Low Temperature Physics}, 15:165, 1989.

\bibitem{Koshino2010}
M.~Koshino and T.~Ando.
\newblock Anomalous orbital magnetism in dirac-electron systems: Role of
  pseudospin paramagnetism.
\newblock {\em Phys. Rev. B}, 81:195431, 2010.

\bibitem{Koshino2016}
M.~Koshino and I.F. Hizbullah.
\newblock Magnetic susceptibility in three-dimensional nodal semimetals.
\newblock {\em Phys. Rev. B}, 93:045201, 2016.

\bibitem{Mikitik2016}
G.~P. Mikitik and Yu.~V. Sharlai.
\newblock Magnetic susceptibility of topological nodal semimetals.
\newblock {\em Phys. Rev. B}, 94:195123, 2016.

\bibitem{Liu2016}
Y.~Liu, Zh. Li, L.~Guo, X.~Chen, Y.~Yuan, F.~Liu, S.~Prucnal, M.~Helm, and Sh.
  Zhou.
\newblock Intrinsic diamagnetism in the {Weyl} semimetal {TaAs}.
\newblock {\em Journal of Magnetism and Magnetic Materials}, 408:73--76, 2016.

\bibitem{Zhang2019}
Ch.-L. Zhang, C.M. Wang, Zh. Yuan, X.~Xu, G.~Wang, Ch.-Ch. Lee, L.~Pi, Ch. Xi,
  H.~Lin, N.~Harrison, H.-Zh. Lu, J.~Zhang, and Sh. Jia.
\newblock Non-saturating quantum magnetization in {Weyl} semimetal {TaAs}.
\newblock {\em Nature Communications}, 10(1):1028, 2019.

\bibitem{Stishov2012}
S.M. Stishov, A.E. Petrova, V.A. Sidorov, and D.~Menzel.
\newblock Self-doping effects in cobalt silicide {CoSi}: Electrical, magnetic,
  elastic, and thermodynamic properties.
\newblock {\em Phys. Rev. B}, 86:064433, 2012.

\bibitem{Narozhnyi2013}
V.~N. Narozhnyi and V.~N. Krasnorussky.
\newblock Studying the magnetic properties of {CoSi} single crystals.
\newblock {\em Journal of Experimental and Theoretical Physics},
  116(5):780--784, 2013.

\bibitem{Shinoda1972}
D.~Shinoda.
\newblock Magnetic properties of {Co$_{1-x}$Fe$_x$Si}, {Co$_{1-x}$Mn$_x$Si},
  and {Fe$_{1-x}$Mn$_x$Si} solid solutions.
\newblock {\em Physica status solidi (a)}, 11(1):129--135, 1972.

\bibitem{Stishov2011}
S.M. Stishov and A.E. Petrova.
\newblock Itinerant helimagnet {MnSi}.
\newblock {\em Phys. Usp.}, 54(11):1117, 2011.

\bibitem{Ishikawa1976}
Y.~Ishikawa, K.~Tajima, D.~Bloch, and M.~Roth.
\newblock Helical spin structure in manganese silicide {MnSi}.
\newblock {\em Solid State Communications}, 19(6):525 -- 528, 1976.

\bibitem{Kanazawa2015}
N.~Kanazawa.
\newblock {\em Charge and Heat Transport Phenomena in Electronic and Spin
  Structures in {B20}-type Compounds}.
\newblock Japan, Springer, 2015.

\bibitem{Gupta2019}
A.~Gupta.
\newblock Novel electric field effects on {Landau} levels in multi-{Weyl}
  semimetals.
\newblock {\em Physics Letters A}, 383(19):2339--2345, 2019.

\bibitem{Pshenay2024}
D.A. Pshenay-Severin, S.N. Nikolaev, Y.V. Ivanov, and A.T. Burkov.
\newblock {Landau} levels of the topological semimetal {CoSi} near the
  {$\Gamma$}-point and their contribution to the orbital magnetic
  susceptibility.
\newblock {\em Physics of the Solid State}, 66(5):665, 2024.

\bibitem{Luttinger1955}
J.~M. Luttinger and W.~Kohn.
\newblock Motion of electrons and holes in perturbed periodic fields.
\newblock {\em Phys. Rev.}, 97:869--883, 1955.

\bibitem{Shoenberg2009}
D.~Shoenberg.
\newblock {\em Magnetic Oscillations in Metals}.
\newblock Cambridge Monographs on Physics. Cambridge University Press, 2009.

\bibitem{McClure1956}
J.~W. McClure.
\newblock Diamagnetism of graphite.
\newblock {\em Phys. Rev.}, 104:666--671, 1956.

\bibitem{Vignale1991}
G.~Vignale.
\newblock Orbital paramagnetism of electrons in a two-dimensional lattice.
\newblock {\em Phys. Rev. Lett.}, 67:358--361, 1991.

\bibitem{Raoux2014}
A.~Raoux, M.~Morigi, J.-N. Fuchs, F.~Pi\'echon, and G.~Montambaux.
\newblock From dia- to paramagnetic orbital susceptibility of massless
  fermions.
\newblock {\em Phys. Rev. Lett.}, 112:026402, 2014.

\end{thebibliography}

\end{document}